\documentclass[12,reprint,showpacs,twocolumn,jcp,nobibnotes]{revtex4} 
\usepackage[hidelinks]{hyperref}
\usepackage[dvipsnames]{xcolor}
\usepackage{graphicx,amsfonts,amsmath,amsbsy,amssymb}
\usepackage[T1]{fontenc}
\usepackage{bm}
\usepackage{amsmath}

\newcommand{\ket}[1]{\left| #1 \right>} 
\newcommand{\bra}[1]{\left< #1 \right|} 

\usepackage{booktabs}
\AtBeginDocument{
  \heavyrulewidth=.08em
  \lightrulewidth=.05em
  \cmidrulewidth=.03em
  \belowrulesep=.65ex
  \belowbottomsep=0pt
  \aboverulesep=.4ex
  \abovetopsep=0pt
  \cmidrulesep=\doublerulesep
  \cmidrulekern=.5em
  \defaultaddspace=.5em
}

\usepackage{glossaries}
\newacronym{hf}{HF}{Hartree-Fock} 
\newacronym{mp2}{MP2}{Moeller-Plesset Perturbation Theory}
\newacronym{CIS}{CIS}{Configuration interaction Singles}
\newacronym{RPA}{RPA}{Random Phase Approximation}
\newacronym{ccsd}{CCSD}{Coupled Cluster Singles Doubles}
\newacronym{uccsd}{UCCSD}{Unrestricted \gls{ccsd}}
\newacronym{ccsd(t)}{CCSD(T)}{perturbative triples} 
\newacronym{dft}{DFT}{Density Functional Theory}
\newacronym{tddft}{TD-DFT}{Time-Dependent Density Functional Theory}
\newacronym{cc}{CC}{Coupled-Cluster}
\newacronym{eom}{EOM}{Equation Of Motion}
\newacronym{eomcc}{EOM-CC}{Equation of motion Coupled-Cluster}
\newacronym{ipeomcc}{IP-EOM-CC}{Ionization Potential \gls{eomcc}}
\newacronym{eaeomcc}{EA-EOM-CC}{Electron Attachment \gls{eomcc}}
\newacronym{eeeomcc}{EE-EOM-CC}{Electron excitation \gls{eomcc}}
\newacronym{eeeomccsd}{EE-EOM-CCSD}{Equation of motion \gls{ccsd}}
\newacronym{cc4s}{\texttt{cc4s}}{Coupled Cluster For Solids}
\newacronym{ctf}{\texttt{CTF}}{Cyclops Tensor Framework}
\newacronym{bse}{BSE}{Bethe-Salpeter equation}

\date{\today}

\begin{document}

\title{%
  A periodic equation-of-motion coupled-cluster implementation
  applied to $F$-centers in alkaline earth oxides
}
\author{Alejandro Gallo}
\author{Felix Hummel}
\author{Andreas Irmler}
\author{Andreas Gr\"uneis}
\affiliation{%
  Institute for Theoretical Physics, TU Wien,\\
  Wiedner Hauptstra{\ss}e 8--10/136, 1040 Vienna, Austria 
}

\begin{abstract}
  We present an implementation of equation of motion coupled-cluster
  singles and doubles (EOM-CCSD) theory using periodic boundary
  conditions and a plane wave basis set.  Our implementation of EOM-CCSD
  theory is applied to study $F$-centers in alkaline earth oxides
  employing a periodic supercell approach.  The convergence of calculated
  electronic excitation energies for neutral color centers in MgO, CaO
  and SrO crystals with respect to orbital basis set and system size is
  explored. We discuss extrapolation techniques that approximate
  excitation energies in the complete basis set limit and reduce finite size errors.
  Our findings demonstrate that EOM-CCSD theory can predict optical
  absorption energies of $F$-centers in good agreement with experiment.
  Furthermore, we discuss calculated emission
  energies corresponding to the decay from triplet to singlet states,
  responsible for the photoluminescence properties.
  Our findings are compared to experimental and theoretical results
  available in literature.
\end{abstract}

\maketitle%

\section{Introduction}

\gls{dft}\cite{Self.ConsistentKohn.1965,Inhomogeneous.EHohenb1964}
using approximate exchange and correlation energy density functionals is
arguably the most successful ab initio approach to compute
materials properties.
Its application goes beyond
ground state properties by providing a reference or starting point
for methods that treat excited-state phenomena explicitly.
In this context, theories such as
\gls{tddft}~\cite{Excitation.EnerPeters1996,Density.FunctioRunge.1984} and
the $GW$ approximation~\cite{New.Method.for.Hedin.1965}
are widely-used to tackle excited states in molecules and
solids~\cite{GW.100.Benchmavan.Se2015,TheGwCompendigolze2019}.
Nonetheless,
they often suffer from a strong dependence on the
\gls{dft} reference calculation.
In the case of \gls{tddft}, albeit
being an exact theory, results depend strongly on the choice of the
approximate exchange and correlation density functional.
Similarly, so-called non-selfconsistent
$G_0W_0$ quasiparticle energies
depend strongly on the Kohn-Sham orbital energies, whereas
fully self-consistent $GW$ calculations are not as often performed
and do not necessarily improve upon the accuracy compared to
$G_0W_0$~\cite{Grumet2018}.
To compute charge neutrality preserving optical absorption energies
from the electron addition and removal energies obtained in the $GW$ framework,
it is necessary to account for the exciton binding energy.
Excitonic effects are often approximated using the
\gls{bse}~\cite{Salpeter1951}.
We note that despite the high level of accuracy and
efficiency of $GW$-\gls{bse} calculations~\cite{Electron.hole.eRohlfi2000},
many choices and approximations have to be made in practice
that are difficult to justify in a pure ab initio framework.
Therefore, it seems worthwhile to explore alternative methods that
are less dependent on \gls{dft} approaches.
\\
\gls{cc}%
~\cite{ Short.range.corCoeste1960%
      , Time.dependent.F..Coe1958%
      , On.the.CorrelatCizek.1966%
      }
formulations are widely-used in the field of molecular quantum chemistry for
both the ground state and excited states via the \gls{eomcc}
formalism~\cite{The.equation.ofStanto1993}.
Ground state \gls{cc} theories such as \gls{ccsd}
and \gls{ccsd(t)}%
~\cite{A.fifth.order.pRaghav1989, Why.CCSD.T.worStanto1997}
have become one of the most successful methods
in molecules in terms of their systematically improvable accuracy and
computational efficiency.
Likewise, \gls{eomcc} methods are routinely applied to molecular
systems with great success%
~\cite{ TheoreticalStuStanto1997%
      , Coupled.clusterBartle2012%
      , piecuchAzulene2013%
      , New.and.EfficieVidal.2019%
      , TowardsARatioIvanov2019%
      }.
However, we stress that the computational cost of \gls{cc} theories is
significantly larger than
that of Green's function based methods mentioned above.
Nonetheless, several studies have focused on making use of these
wavefunction methods also in solids
to study ground and excited state
properties%
~\cite{ Applying.the.CoGruber2018%
      , Electronic.struGao.Y2020%
      , Gaussian.based.James.2017%
      }.
While \gls{eom} type methods are well understood and benchmarked in
finite systems,
this is less so for periodic systems, where ongoing efforts
are made towards applications in solids.
Previous applications of \gls{eom} type methods have focused on electronic
band structures
using the \gls{ipeomcc} and \gls{eaeomcc}
extensions%
~\cite{ Gaussian.based.James.2017%
      , Electronic.struGao.Y2020%
      , First.principlePulkin2020%
      , Electronic.struGao.Y2020%
      , Spectral.functiMcClai2016%
      }
as well as its \gls{eeeomcc}
extensions~\cite{Equation.of.motKatagi2005,Excitons.in.SolWang.2020},
all of which are based on Gaussian basis sets.
For local phenomena, such as defect excitation energies, several
studies have been performed employing cluster models of the periodic
structures~\cite{Ab.initio.perspTiwald2015,OnTheAccuratesousa2001}.
One of the main challenges in
these calculations is to achieve a good control over the finite basis
set and system size errors, which is often achieved using
extrapolation techniques.
In this manuscript, we study excited state properties of
point defects in solids, computed on the level of \gls{eeeomcc}.
Understanding impurities in solids is important for both theoretical
and practical reasons. Lattice defects affect bulk properties
of the host crystal and both the understanding of ground
and excited-state properties is essential for these
systems~\cite{Structure.des.cFriede1967, Theory.of.electVail.1990}.
Here, we focus on color centers in the alkaline
earth oxide crystals MgO, CaO and SrO in the rock salt structure.
Removing an oxygen atom from these systems
results in so-called $F$-centers that can be filled by
2 ($F^0$), 1 ($F^+$) or 0 ($F^{2+}$) electrons.
The corresponding one-electron states are stabilized by the Madelung potential
of the crystal and their electron density is in general localized in the cavity
formed by the oxygen vacancy.
These defects are typically produced by neutron
irradiation~\cite{LuminescenceOfrosenb1989}
or additive colorization~\cite{Photoluminescenedel1979}.
Much effort has been made to elucidate the exact mechanism
of the luminescence of $F$-centers in MgO, CaO and SrO
\cite{OnTheLuminescbartra1975, The.theory.of.dStoneh1979}.
The ground and excited state properties of these vacancies are
of importance for a wide range of technological applications
including color center lasers.
Furthermore, vacancies of oxides are of general importance for understanding
their surface chemistry and related properties.
In this work we will concentrate on the diamagnetic $F^0$-center.
The trapped electrons can be viewed as a pseudo-atom embedded in a solid,
where the optical absorption and emission between ground and low-lying
excited states is characterized by the electron transfer between
1$s$ into 2$s$ or 2$p$ one-electron states.
Initial theoretical studies of these defects were already performed in
the 1960s and 1970s using effective Hamiltonians%
~\cite{WaveFunctionsKemp1963, ElectronicStruwood1975, ElectronicStruwilson1977}.
Modern ab initio studies of the $F^0$ center in MgO
have employed cluster approaches in combination with quantum chemical
wavefunction based methods~\cite{OnTheAccuratesousa2001}, fully periodic
supercell approaches in combination with the \(GW\)-\gls{bse}
approach~\cite{First.PrincipleRinke.2012,PredictionOfOtosoni2012} or Quantum
Monte Carlo calculations~\cite{PointDefectOperteki2013}.
In this work we seek to employ a periodic supercell approach and
a novel implementation of \gls{eeeomccsd} theory using a plane wave
basis set.
In addition to the $F$-center in MgO, we will also study $F$-centers
in CaO and SrO.  We note that \gls{eeeomccsd} theory is exact for
ground and excited states of two electron systems and is therefore
expected to yield very
accurate results for the F$^0$ center in alkaline earth oxides.
We will discuss different techniques to correct for
finite basis set and supercell size errors and demonstrate that
\gls{eeeomccsd} theory can be used to compute accurate absorption and
emission energies compared to experiment without the need for
adjustable parameters and the ambiguity caused by the choice of the starting
point.

The following is a summary of the structure of this work.
In Section~\ref{sec:methods} we give a brief overview of the
employed theoretical and computational methods used to compute
excitation energies including extrapolation
techniques that are needed to approximate to the complete basis set
and infinite system size limit.
Section~\ref{sec:results} presents the obtained results of the defect
calculations and draws a comparison between this work and available
experimental and theoretical results from the literature.

\section{Theory and Methods}\label{sec:methods}

We start this section by giving a brief description of the employed
\gls{cc} methods followed by a discussion of the computational
details.

\subsection{\gls{ccsd} theory}

The \gls{cc} approximation is based on an exponential \textit{ansatz} for
the electronic
wavefunction~\cite{On.the.CorrelatCizek.1966,Coupled.clusterBartle2007}
acting on a single Slater determinant $\ket{0}$,
\[
\ket{\Psi}_{\mathrm{CC}} = e^{\hat{T}} \ket{0}
\]
where the \textit{cluster operator} consists of second-quantized
neutral excitation operators
\[
  \hat{T} = \sum_{\mu} t_{\mu} \hat{\tau_{\mu}}
  ,\qquad
  t_{\mu} \in \mathbb{C}
\]
with $\mu$ labeling excitation configurations. For instance,
when considering only singles and doubles excitations (\gls{ccsd})
the unrestricted \gls{ccsd} cluster operator is given by
\begin{equation*}
\hat{T}
  = \sum_{a,i}t^a_i \hat{a}^{\dagger}_{a}\hat{a}_{i}
  + \sum_{a,b,i,j}
      t^{ab}_{ij}
      \hat{a}^{\dagger}_{a} \hat{a}^{\dagger}_{b}
      \hat{a}_{j}         \hat{a}_{i}
\end{equation*}
where the set of indices \( \{a,b,c, \ldots \} \)
denote virtual or unoccupied spin orbitals and \( \{ i,j,k, \ldots \} \)
denote occupied spin orbitals.
Orbitals are occupied or unoccupied with respect to the
reference Slater determinant $\ket{0}$, which may come for instance
from a \gls{hf} or a \gls{dft} calculation.
Here, we will restrict the discussion to the case of \gls{ccsd}.
Applying the \gls{cc} \textit{ansatz} to the stationary many-body electronic
Schr\"odinger equation results in
\begin{equation}
\bar{H} \ket{0}
  = e^{-\hat{T}}\hat{H}e^{\hat{T}} \ket{0}
  = E_{\mathrm{CC}} \ket{0}
\end{equation}
where \( E_{\mathrm{CC}} \) is the coupled cluster energy, and we have
implicitly defined the similarity transformed Hamiltonian \( \bar{H} \).
The state \( \ket{\Psi_{\mathrm{CC}}} \)
is parametrized by the coefficients \( t_{\mu} \),
which can be obtained by projection.
In the case of \gls{ccsd} one projects the
Schr\"odinger equation onto the singles and doubles sections of
the \textit{Fock} space
\begin{align}
  \label{eq:ccsd-equations-energy}
  E_{\mathrm{CC}} &= \bra{0} \bar{H} \ket{0} \\
  \label{eq:ccsd-equations-singles}
  0  &= \bra{0}
          \hat{a}^{\dagger}_{i}\hat{a}_{a} \bar{H}
        \ket{0} \\
  \label{eq:ccsd-equations-doubles}
  0  &= \bra{0}
          \hat{a}^{\dagger}_{i}\hat{a}^{\dagger}_{j}
          \hat{a}_{b} \hat{a}_{a}
        \bar{H}
        \ket{0}
  .
\end{align}
Equations~(\ref{eq:ccsd-equations-energy}--\ref{eq:ccsd-equations-doubles})
are a set of coupled non-linear equations
in terms of the amplitudes \( t^{a}_{i} \) and \( t^{ab}_{ij} \)
that are solved by iterative methods.

\subsection{\gls{eeeomccsd} theory}

A common way to obtain excited states based on the \gls{cc} theory
is through diagonalizing the similarity transformed Hamiltonian
\( \bar{H} \) in a suitable subspace of
the Fock space~\cite{The.equation.ofStanto1993}.
We are going to present the neutral variant of this approach, also called
electronically excited equation of motion, for which the number of electrons
is conserved.
In consequence, restricting from now on again the analysis to singles
and doubles excitations, the \textit{ansatz} for an excited state
$\hat{R}\ket{\Psi_{\mathrm{CC}}}$
is
\begin{equation}
  \label{eq:main-eom-ansatz}
    \hat{H} \hat{R}\ket{\Psi_{\mathrm{CC}}}
  = \hat{H} \hat{R} e^{\hat{T}} \ket{0}
  = E_{R}   \hat{R}\ket{\Psi_{\mathrm{CC}}}
\end{equation}
where
\begin{equation}
\hat{R} = r_{0}
        + \sum_{a,i} r^a_i \hat{a}^{\dagger}_{a}\hat{a}_{i}
        + \sum_{a,b,i,j}
            r^{ab}_{ij}
            \hat{a}^{\dagger}_{a} \hat{a}^{\dagger}_{b}
            \hat{a}_{j}         \hat{a}_{i}
        ,\qquad r_{\mu} \in \mathbb{C}
\end{equation}
is a linear excitation operator and \( E_{R} \) is its excitation energy.
Equation~\eqref{eq:main-eom-ansatz} is equivalent to a commutator equation
only involving \( \bar{H} \) and the excitation energy difference
\( \Delta E_{R} \) between \( E_R \) and the correlated ground state
\( E_{\textrm{CC}} \),
\begin{equation}
  \label{eq:eom-commutator-relation}
    [\bar{H}, \hat{R}] \ket{0}
  = {\left(
      \bar{H}
      \hat{R}
    \right)}_{\mathrm{c}} \ket{0}
  = (E_{R} - E_{\mathrm{CC}}) \hat{R} \ket{0}
  = \Delta E_{R} \hat{R} \ket{0}
  .
\end{equation}
It is worthwhile noting that the commutator on the left-hand-side
means that only connected diagrams need to be considered in the
CI expansion, which is denoted by the parentheses \( {()}_{\mathrm{c}} \).
Equation~\eqref{eq:eom-commutator-relation} motivates the name
\textit{equation of motion} due to its resemblance to the time-dependent
\textit{Heisenberg picture} differential equation for the time evolution
of an operator.

\subsection{Computational methods and details}

\begin{figure}[t]
  \includegraphics[width=0.35\textwidth]{%
    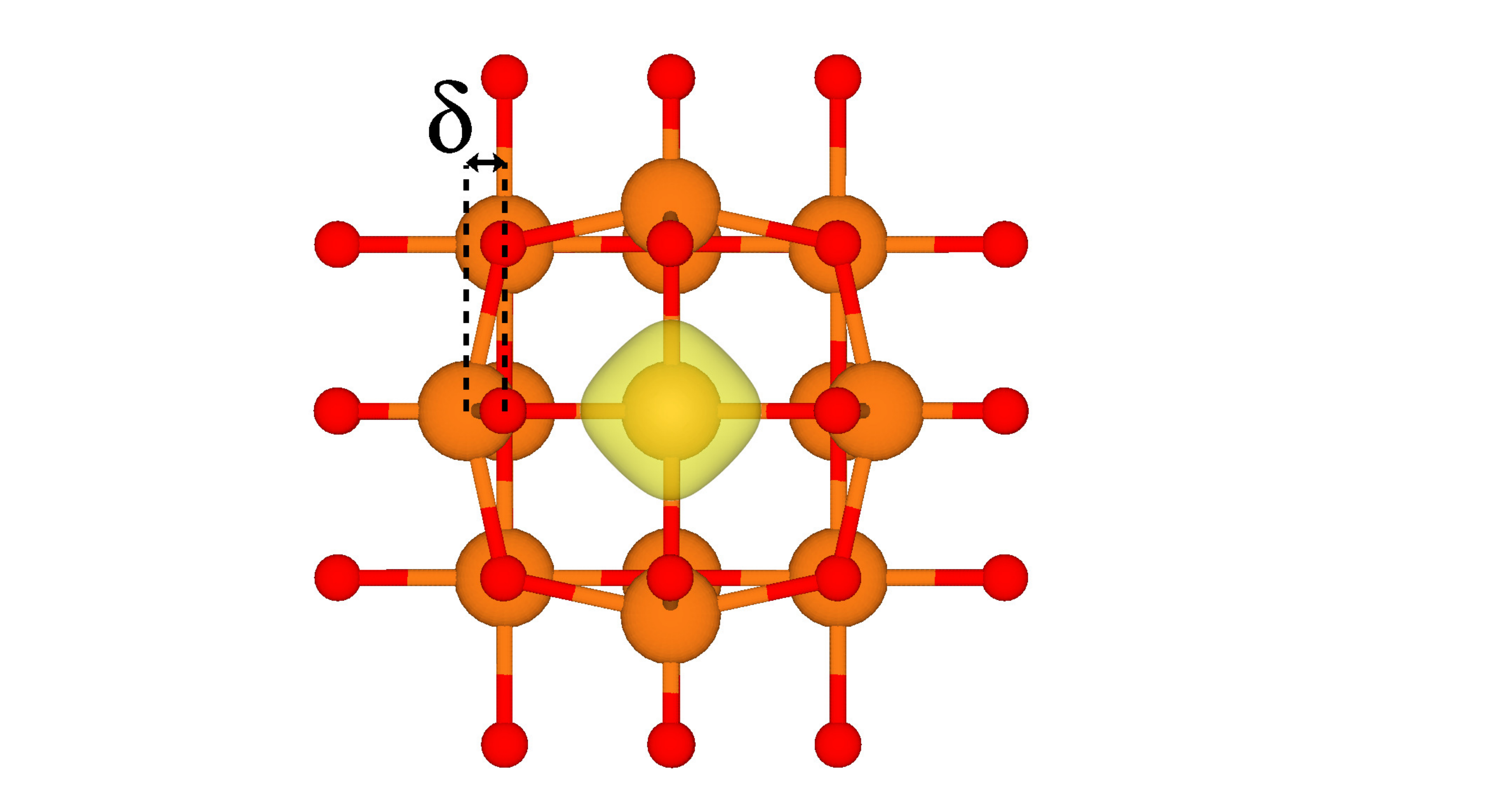}
  \caption{%
    Geometry of neutral $F$-center in MgO. Red and orange spheres
    correspond to oxygen and magnesium atoms, respectively. The yellow
    isosurface was computed from the localized electronic states in the
    band gap of MgO that originates from the two trapped
    electrons. $\delta$ measures the displacement along the $A_{1g}$
    vibrational mode with the Mg atoms out of
    their equilibrium position in the bulk structure and was deliberately
    chosen larger for this figure to emphasize the effect of lattice
    relaxation.
  }
  \label{fig:f_in_mgo}
\end{figure}

Here, all \gls{eeeomccsd} calculations of defective supercells employ
a \gls{hf} reference.  The \gls{hf} calculations are performed using the
Vienna ab initio simulation package
(\texttt{VASP})\cite{Efficiency.of.aKresse1996}
and a plane wave basis set in the framework of the projector augmented wave
(PAW)\cite{Projector.augmeBlochl1994} method.
The energy cutoff for the plane wave basis set is 900~eV.
The defect geometries have been relaxed on the level of
\gls{dft}-PBE, starting from a defective geometry with
the corresponding equilibrium lattice constant
(MgO: 4.257~\AA, CaO: 4.831~\AA, SrO: 5.195~\AA)
keeping the lattice vectors and volume fixed.
In this work we study defective 2$\times$2$\times$2,
3$\times$3$\times$3 and 4$\times$4$\times$4 fcc supercells containing
15, 53, and 127 atoms, respectively.
The oxygen vacancy results in an outward relaxation
of the alkaline earth atoms away from the cavity created by the oxygen vacancy.
This outward relaxation strongly overlaps with the vibrational mode $A_{1g}$
and is illustrated in Fig.~\ref{fig:f_in_mgo}.
While the \gls{dft}-PBE calculations have been carefully checked for
convergence with respect to the
$k$-point mesh used to sample the first Brillouin zone, all \gls{hf} and post-\gls{hf}
calculations employ the $\Gamma$-point approximation.

We have implemented \gls{uccsd} and \gls{eeeomccsd} in the \gls{cc4s} code that
was previously employed for the study of various ground state
properties of periodic
systems~\cite{Applying.the.CoGruber2018,Low.rank.factorHummel2017}.
The employed Coulomb integrals and related quantities
were calculated in a completely analogue manner.
Our \gls{uccsd} implementation
is based on the intermediate amplitudes approach
of Stanton et al.~\cite{A.direct.producStanto1991}.
On the other hand, our \gls{eeeomccsd} implementation
uses intermediates for the similarity transformed Hamiltonian
$e^{-\hat{T}}\hat{H}e^{\hat{T}}$
from Stanton et al.~\cite{The.equation.ofStanto1993}
and Shavitt et al.~\cite{ManyBodyMetShavit2009}.
We use the \gls{ctf}\cite{A.massively.parSolomo2014} for the implemented
computer code, which enables an automated parallelization of the underlying
tensor contractions.\\

The diagonalization of the similarity transformed Hamiltonian
is done using a generalized
Davidson solver\cite{A.generalizatioHirao.1982,A.Comparison.ofCarica2010}
which enables the calculation of \gls{eeeomccsd}
energies without explicit calculation of the left eigenvectors.
For the initial guess of the eigenvectors, we use the one-body
\gls{hf} excitation energies and corresponding Slater determinants.
The \gls{uccsd} and \gls{eeeomccsd} calculations have been performed
using only a small number of active \gls{hf} orbitals around the Fermi energy
of the employed supercells. Most occupied orbitals at low energies
are frozen and the same applies to all unoccupied orbitals above a
certain cutoff energy.
The following sections summarize the benchmarks of the implemented
\gls{eeeomccsd} code and investigate the convergence behavior of
the computed excitation energies with respect to the number of active
orbitals as well as system size.

\subsubsection{Benchmark results}

In the following we discuss benchmark results of our \gls{eeeomccsd}
implementation and outline our approach to identify the spin multiplicity
attributed to the excited states.
To verify the implemented expressions,
we have compared the computed \gls{eeeomccsd} excitation energies to
results computed using a well-established quantum chemical code
\texttt{NWCHEM}\cite{NwchemPastPapra2020}.  As most quantum chemical
codes (including \texttt{NWCHEM}) employ atom-centered Gaussian basis
sets,
it was also necessary to implement an interface that reads the
orbital coefficients from \texttt{NWCHEM} and
employs the \texttt{LIBINT2}\cite{Libint2} library to compute corresponding
integrals.
As test systems we have selected the neon atom and the water molecule
in the \emph{aug-cc-pvdz} basis.
All computed \gls{uccsd} energies using \gls{hf} and \gls{dft} reference
determinants
achieve an excellent agreement (8 significant digits) between both codes.
For \gls{eeeomccsd} calculations, the singlet states computed by \texttt{NWCHEM}
were also obtained using our \gls{eeeomccsd} implementation with similar
convergence behavior and in excellent agreement (8 significant digits).
We can identify the triplet and the singlet state in our output
by using spin-flip \gls{eeeomccsd}~\cite{Equation.of.MotKrylov2008}
and comparing the degeneracy of the states computed with and without
spin-flip excitations.
In future work we will implement the direct computation
of the spin expectation value.

\subsubsection{Orbital basis convergence of excited states}

\begin{figure}[t]
  \includegraphics[width=0.45\textwidth]{%
    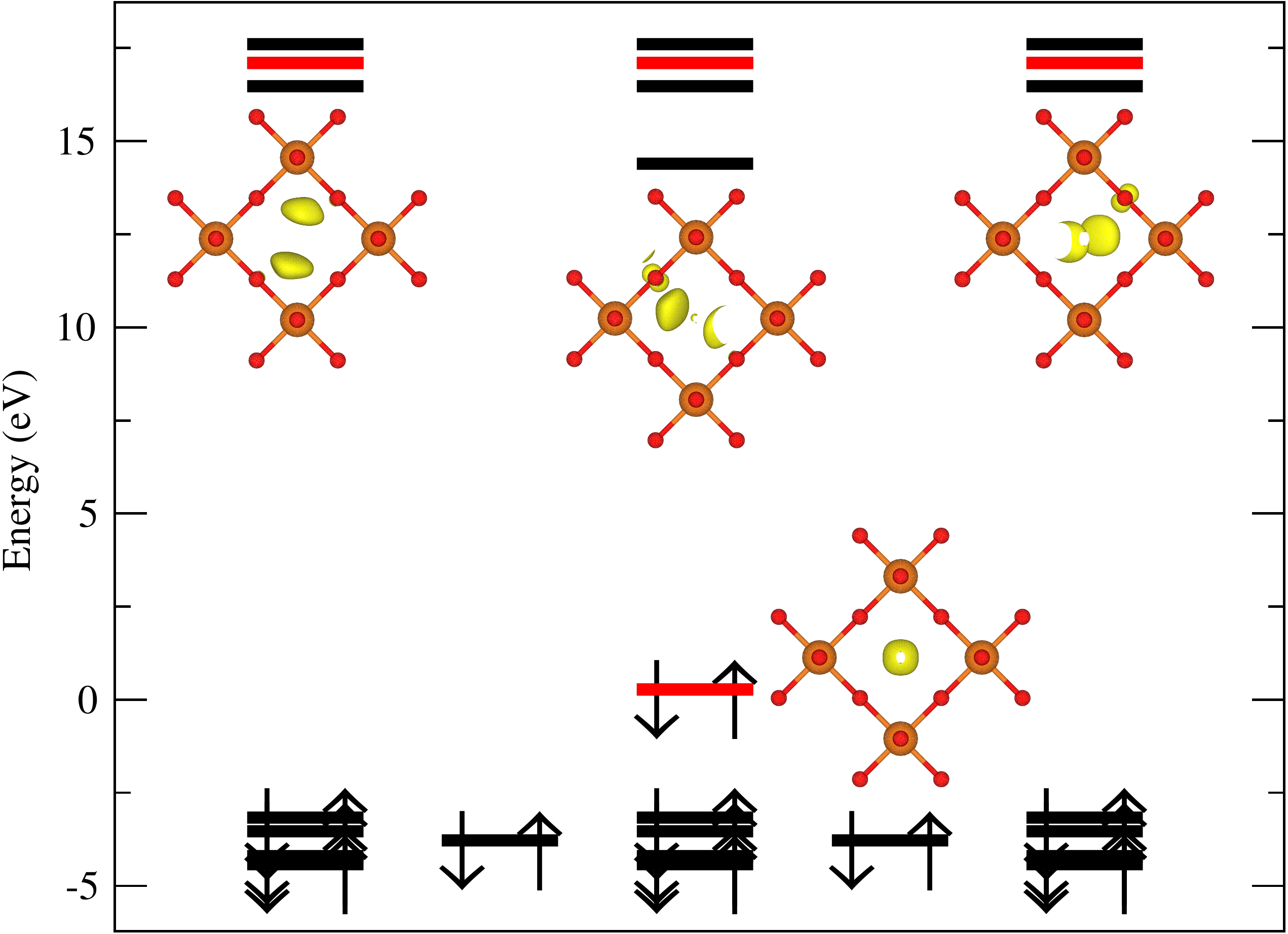}
  \caption{%
    Occupied and virtual \gls{hf} energy levels.  The red levels
    correspond to defect states and the corresponding isosurfaces of the
    charge densities are depicted.
  }
  \label{fig:f_in_mgo_levels}
\end{figure}

All presented findings in this section have been obtained for the $F$-center
in MgO. However, the corresponding findings for CaO and SrO are qualitatively
identical unless stated explicitly.

We first seek to investigate
the character of the employed \gls{hf} orbitals
and the convergence of the computed excitation
energies with respect to the canonical orbital basis set size.
The \gls{hf} orbitals have been computed for a defective 2$\times$2$\times$2
MgO supercell containing 15 atoms.
Figure~\ref{fig:f_in_mgo_levels} depicts the energy levels around the
Fermi energy and isosurfaces of charge densities computed for the
defect states.  The occupied state with the highest one-electron
energy corresponds to the occupied defect state and
its orbital energy is located in the gap of the bulk crystal.
Its charge density is well localized
in the cavity created by the oxygen vacancy.
In the thermodynamic limit (big supercells or dense
$k$-meshes), the direct and fundamental gap of pristine MgO is 15.5~eV
on the level of \gls{hf} theory~\cite{Second.order.MoGrunei2010}, which is
significantly larger
than the experimental gap of about 7.8~eV.  The neglect of correlation
effects in \gls{hf} theory overestimates band gaps for a wide range of
simple semiconductors and insulators.
The orbital ordering between defect and bulk states
depicted in Fig.~\ref{fig:f_in_mgo_levels} is
qualitatively identical to the one observed for CaO and SrO.
However, we stress that in contrast to MgO, CaO and SrO exhibit an indirect
band gap with a conduction band minimum at the Brillouin zone boundary.

We note that the supercells investigated in this work contain up to 127 atoms,
corresponding to more than 1000 valence electrons.
The computational cost of \gls{eeeomccsd} theory scales as
$\mathcal{O}(N^6)$, where $N$ is some measure of the system size.
In particular, the cost for some of the most important tensor
algebraic operations scales as
$\mathcal{O}(N_v^4N_o^2)$
and
$\mathcal{O}(N_v^2N_o^4)$,
where $N_o$
and $N_v$ refer to the number of occupied and virtual orbitals,
respectively.
Additionally, the memory
footprint of our implementation scales as $\mathcal{O}(N^4)$.
Due to the steep scaling of the computational cost, an explicit
treatment of all electrons on the level of \gls{eeeomccsd} becomes
intractable and renders it necessary to freeze a large fraction of the
occupied and virtual \gls{hf} states.  In the following we will
investigate the convergence of the computed excitation energies with
respect to the number of active virtual and occupied states.

\begin{figure}[t]
  \includegraphics[width=0.5\textwidth]{%
    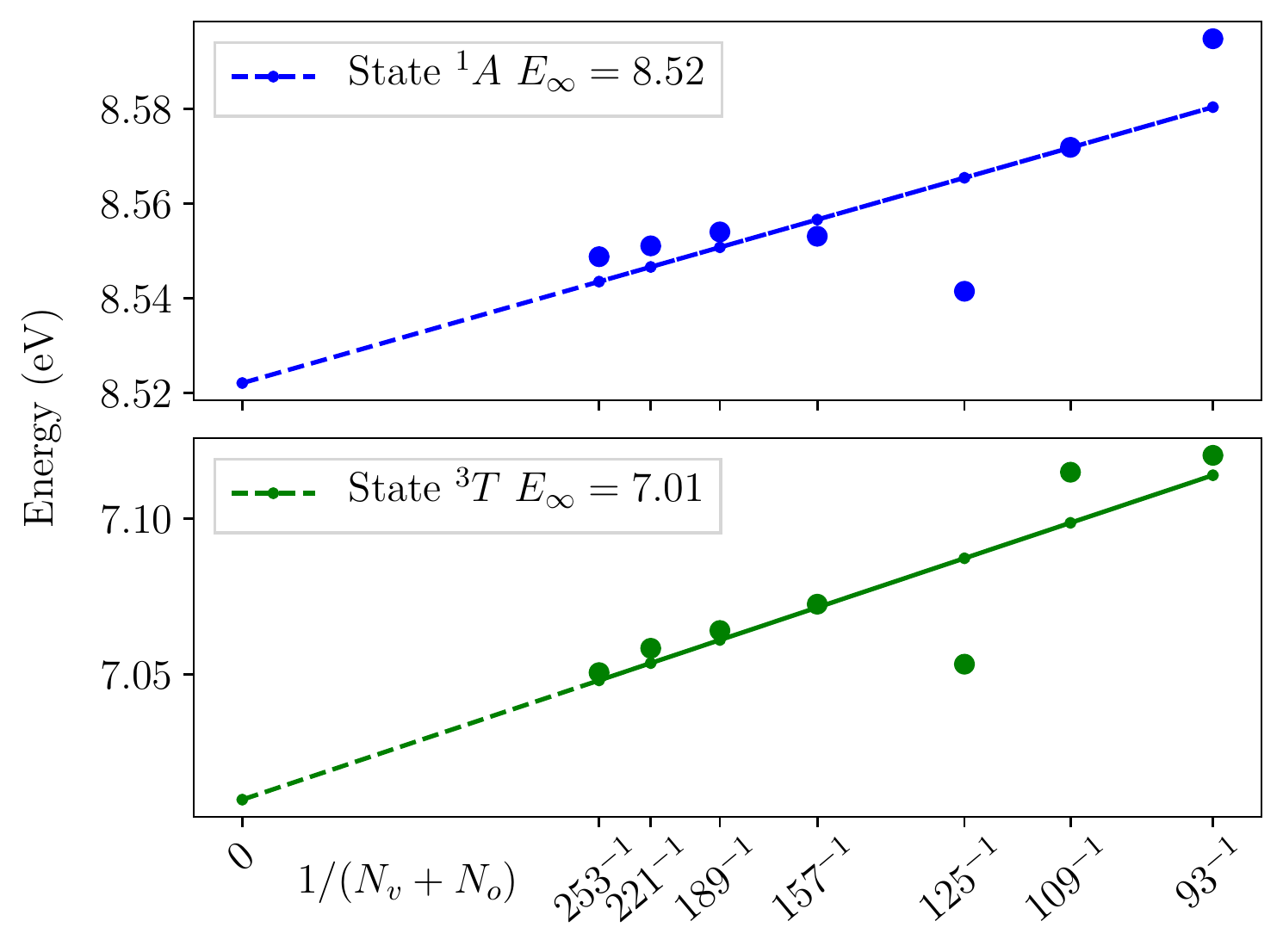}
  \caption{%
    Basis set extrapolation of lowest \gls{eeeomccsd} excitation energies
    corresponding to excitations of the $F$-center defect in MgO.
    All computed energies have been fitted against
    $1/(N_v + N_o)$, where $N_v$ and $N_o$ is the number of virtual
    and occupied orbitals used.
    The lower and higher excitation energies correspond to a
    singlet-triplet and a singlet-singlet transition, respectively.
    This extrapolation has been obtained for a supercell composed of
    eight Mg and seven O atoms.
  }
  \label{fig:mgo-basis-set-extrapolation222}
\end{figure}

We first investigate the convergence of \gls{eeeomccsd} excitation
energies with respect to the virtual orbital basis set.
Among the 61 occupied spatial \gls{hf} orbitals we keep only
the four orbitals active with the highest energy.
Furthermore, we only investigate many-electron excited states with
$r^a_i$ excitation amplitudes that correspond to a significant charge
transfer from the occupied $s$-like defect state to the virtual $p$-like
defect states as illustrated in Fig.~\ref{fig:f_in_mgo_levels}.
Fig.~\ref{fig:mgo-basis-set-extrapolation222} depicts the convergence of
the \gls{eeeomccsd} excitation energies that we assign to local
excitations of the $F$-center.
In passing we note that \gls{eeeomccsd} theory predicts a number
of excited states that describe electronic excitations
with charge transfer from the defect to bulk states, which
will not be explored in this work.
The electronic ground state of the neutral $F$-centers
studied in this work is a singlet state.
The lower and higher excitation energies shown in
Fig.~\ref{fig:mgo-basis-set-extrapolation222}
correspond to a singlet-triplet and singlet-singlet transition energy,
respectively.  We observe for both excitation energies a
$1/(N_{v} + N_{o})$
convergence to the complete basis set limit.  This behavior is not
unexpected and agrees with the convergence of ground state energies.
Furthermore, we note that a similar convergence was observed for
\gls{eeeomccsd} exciton energies of bulk
materials~\cite{Excitons.in.SolWang.2020}.
We note that it might seem advantageous to replace \gls{hf} virtual orbitals
with a different type of orbitals; for example, natural orbitals, to
accelerate the convergence.  However, we have found that these
orbitals will mostly accelerate the convergence of the ground state
energy, introducing large basis set incompleteness errors in the
convergence of excitation energies.  In this work we will employ a
$1/(N_{v} + N_{o})$ extrapolation to approximate excitation energies
in the complete basis set limit of all systems.

\begin{figure}[t]
  \includegraphics[width=0.5\textwidth]{%
    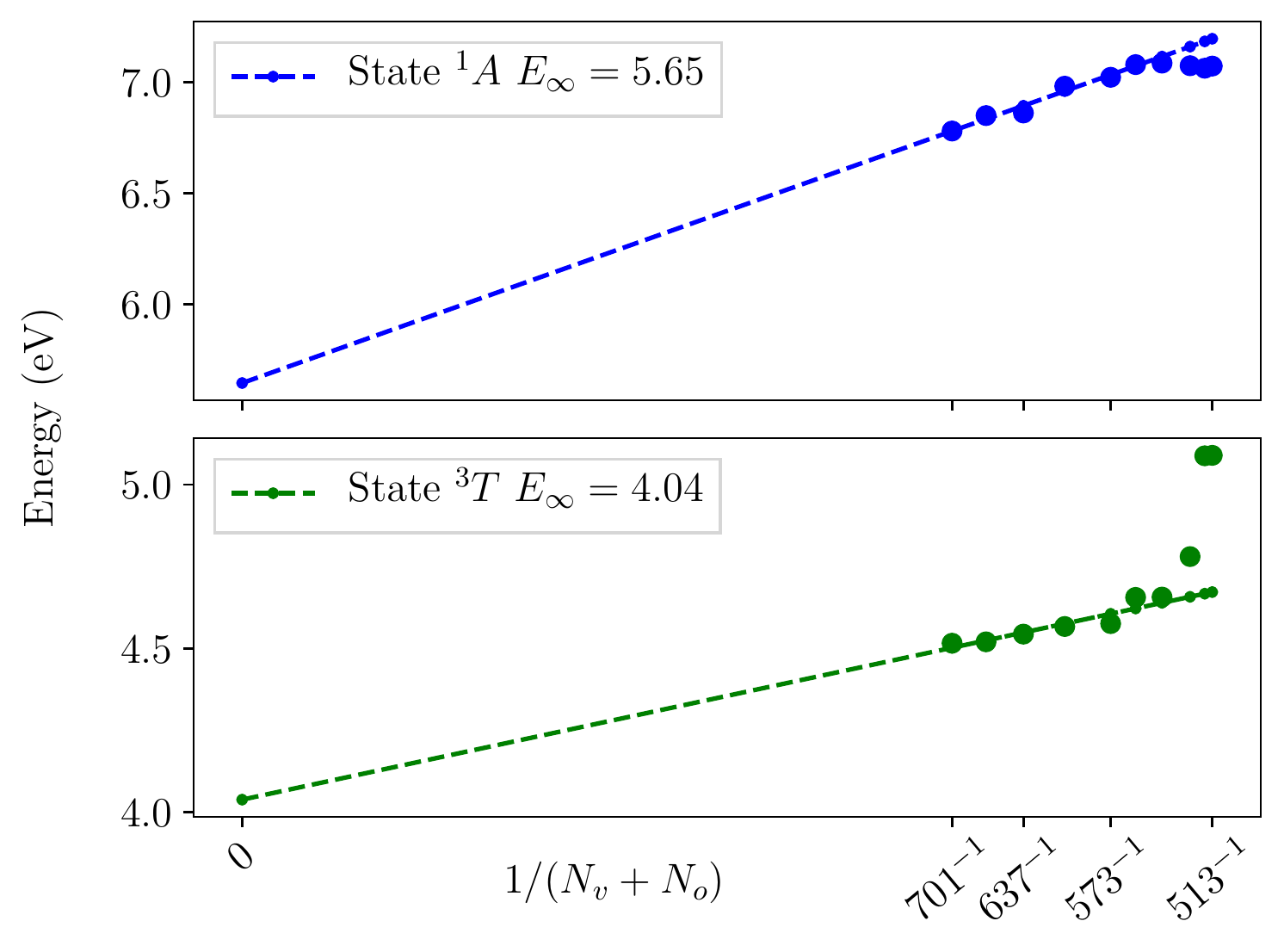}
  \caption{%
    Basis set extrapolation of lowest \gls{eeeomccsd} excitation energies
    corresponding to excitations of the $F$-center defect in MgO
    using a 4$\times$4$\times$4 supercell.
    The fit has been performed ignoring the first four data points.
    States, energies and fit are to be interpreted as in
    Fig.~\ref{fig:mgo-basis-set-extrapolation222}.
  }
  \label{fig:mgo-basis-set-extrapolation444}
\end{figure}

Fig.~\ref{fig:mgo-basis-set-extrapolation444} shows the employed
basis set extrapolation for identical transitions in
a larger 4$\times$4$\times$4 supercell. We note that the slope
of the excitation energy extrapolation is significantly steeper compared
to the 2$\times$2$\times$2 supercell
shown in Fig.~\ref{fig:mgo-basis-set-extrapolation222}.
This can be attributed to the smaller number of virtual orbitals
relative to the complete basis set size for the given plane wave cutoff energy.
Therefore we ignore the first 4 points in the extrapolation
for \emph{all} systems in the 4$\times$4$\times$4 supercell.
In the case of CaO and SrO, the basis set convergence of the excitation energies
is qualitatively identical, and we employ the same orbital basis set sizes in
all extrapolations.

\begin{figure}[t]
  \includegraphics[width=0.4\textwidth]{%
    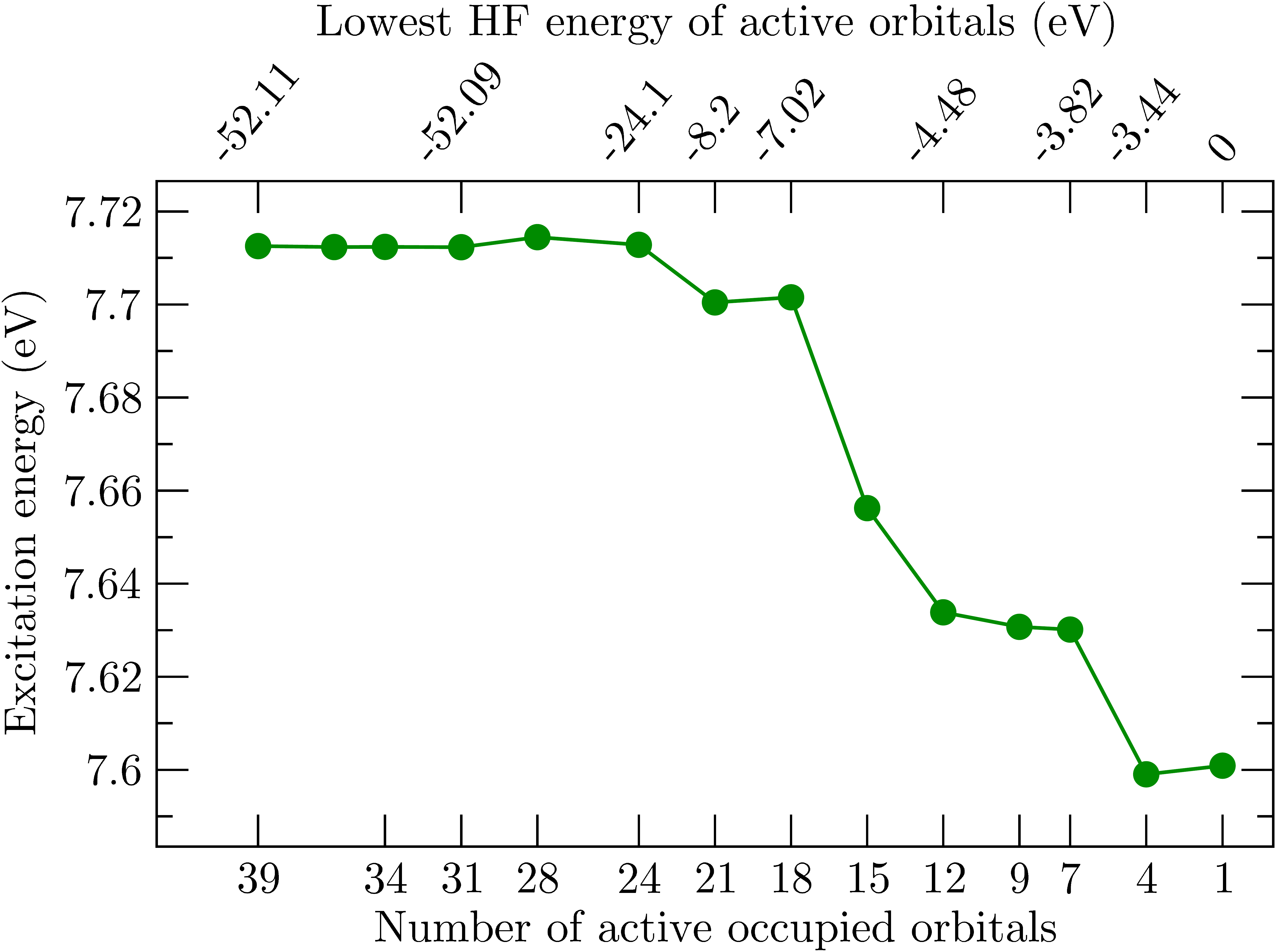}
  \caption{%
    Convergence of the \gls{eeeomccsd} excitation energy for the
    singlet-triplet transition in the $F$-center of MgO with respect to the
    number of inactive/frozen occupied orbitals in
    the \gls{eeeomccsd}  calculation.
    For the employed supercell the \gls{hf}, calculations have been performed
    using 61 occupied and 10 virtual orbitals. The top horizontal axis
    shows the lowest \gls{hf} energy of the included active
    occupied orbital relative to the occupied defect state. All
    orbitals with a lower energy have not been included in the respective
    \gls{eeeomccsd} calculation.
  }
  \label{fig:occ_conv}
\end{figure}

We now investigate the convergence of the \gls{eeeomccsd} excitation
energies with respect to the number of active occupied orbitals,
keeping a virtual orbital basis set consisting of 10 unoccupied
orbitals and employing a 2$\times$2$\times$2 supercell only.
Figure~\ref{fig:occ_conv} depicts the convergence of the lowest defect
excitation energy (singlet-triplet transition) with respect to the
size of the active occupied orbital space.
The horizontal axis at the bottom shows the number of active occupied
orbitals. The horizontal axis at the top of Fig.~\ref{fig:occ_conv} shows
the corresponding lowest \gls{hf} orbital energy.
Our findings demonstrate that the excitation energy increases with
respect to the number of active occupied orbitals and is well
converged to within a few meV using more than about 25 occupied orbitals.
However, a comparison between the converged result and a minimal
active occupied orbital space,
consisting of the occupied defect orbital only, reveals that such a
truncation introduces excitation energy errors of about 120~meV.
We note that one-electron states with relative energies below $-50$~eV
exhibit Mg 2$p$ and 2$s$ character and are therefore expected to be
negligible for the computed excitation energies.
From the above findings we conclude that the excitations studied in
the present work exhibit a significantly larger error from the virtual
orbital basis truncation than from the occupied orbital basis truncation.
Due to the computational cost of \gls{eeeomccsd} calculations we will
therefore extrapolate the excitation energy to the complete basis set
limit while using only 4 occupied orbitals.

\subsubsection{System size convergence of excitation energies}
\label{sec:sysconv}

\begin{table}[t]
  \caption{%
    Convergence of the $F$-center's excitation
    energies in MgO, CaO and SrO for increasing supercell size.  TDL
    corresponds to the extrapolated thermodynamic limit estimate of the
    respective excitation energies assuming a $1/N$ convergence and
    employing the energies of the 2$\times$2$\times$2 and 4$\times$4$\times$4
    supercells. Here $N$ stands for a measure of the system size.
    In this case, the number of electrons is used.
    All energies in eV units.
  }
  \label{tab:tdl-conv}
  \begin{tabular}{llrr}
    \toprule
    System & Supercell       & $^3T$   & $^1A$  \\
    \midrule
    MgO & 2$\times$2$\times$2  & 7.009   & 8.522  \\
        & 3$\times$3$\times$3  & 4.866   & 6.571  \\
        & 4$\times$4$\times$4  & 4.038   & 5.646  \\
        & TDL                  & 3.660   & 5.281  \\
    \midrule
    CaO & 2$\times$2$\times$2  & 3.224  & 3.338  \\
        & 3$\times$3$\times$3  & 2.951  & 4.025  \\
        & 4$\times$4$\times$4  & 2.081  & 3.157  \\
        & TDL                  & 1.936  & 3.134  \\
    \midrule
    SrO & 2$\times$2$\times$2  & 2.324  & 2.413  \\
        & 3$\times$3$\times$3  & 2.404  & 3.155  \\
        & 4$\times$4$\times$4  & 1.332  & 2.351  \\
        & TDL                  & 1.206  & 2.343  \\
    \bottomrule
  \end{tabular}
\end{table}

Having discussed basis set convergence of the computed \gls{eeeomccsd}
excitation energies, we now turn to the discussion of their convergence
with respect to supercell size.  Excitation energies are intensive
quantities. However, their convergence with respect to
system size can sometimes be extraordinarily slow.
We have computed the $F$-center's singlet-triplet and singlet-singlet
transition energies for three different supercell sizes containing 15,
53 and 127 atoms.  Table~\ref{tab:tdl-conv} lists the
computed excitation energies for all systems using different
supercell sizes.
The excitation energies have been obtained using 4 active
occupied orbitals only and extrapolating to the complete
basis set limit as discussed in the previous sections.

We note that
the excitation energies converge monotonously for MgO
with increasing supercell size, but
show some non-monotonic behaviour for the other two systems studied.
This can be explained by the fact that CaO and SrO exhibit a
conduction band minimum at the Brillouin zone boundary.
The electronic states at the conduction band minimum are
therefore only accounted for when using supercells that are constructed
from even-numbered multiples of the fcc unit cell.
Neglecting these important states around the Fermi energy leads to
a significant overestimation of the excitation energies for the excited
singlet states as can be seen by comparing the results obtained for
the 3$\times$3$\times$3 supercell to findings for the 2$\times$2$\times$2
and 4$\times$4$\times$4 supercells.

Here, we seek to remove the remaining finite size errors of the
excitation energies by performing an extrapolation to the
infinite system size limit assuming a 1/$N$ convergence, where $N$ is the total
number of electrons in each supercell.
This approach is in agreement with procedures that are applied to
ground state energy calculations~\cite{Applying.the.CoGruber2018,Liao2016}.
For the sake of consistency we employ only
2$\times$2$\times$2 and 4$\times$4$\times$4 supercells
for the extrapolation for all three studied systems.

Our findings show that the excitation energies
decrease significantly with increasing supercell size in the case of MgO.
Changing the supercell size from a 2$\times$2$\times$2 to a
4$\times$4$\times$4 cell results in a lowering of the excitation
energies by almost 3~eV. This relatively slow
convergence is expected to originate from
strongly delocalized excited defect states of the neutral
$F$-center in MgO.
We note in passing that the excitation energies of the $F$-centers in
CaO and SrO exhibit a significantly faster convergence with respect
to system size.
We attribute this behavior to a more localized character of the
excited $F$-center in CaO and SrO compared to MgO that might be explained
by the significantly smaller size of the cavity formed by the oxygen
vacancy in MgO compared to CaO or SrO.

\section{Results}\label{sec:results}

In this section we describe
the photochemical process of absorption
and emission in the $F$-center of alkaline earth oxides.
We first discuss the energies of the
electronically excited defect states
as a function of the atomic displacements along
the $A_{1g}$ vibrational mode in MgO
to introduce the emission model.
Next, we present our results for the absorption and emission of the $F$-center
in MgO, where problems in the interpretation of the
experimentally observed luminescence band are discussed additionally.
We end this section with a discussion of
the results for CaO and SrO.

\subsection{Absorption and emission process in $F$-centers}
\label{sec:confcoord}

\begin{figure}[t]
  \includegraphics[width=0.5\textwidth]{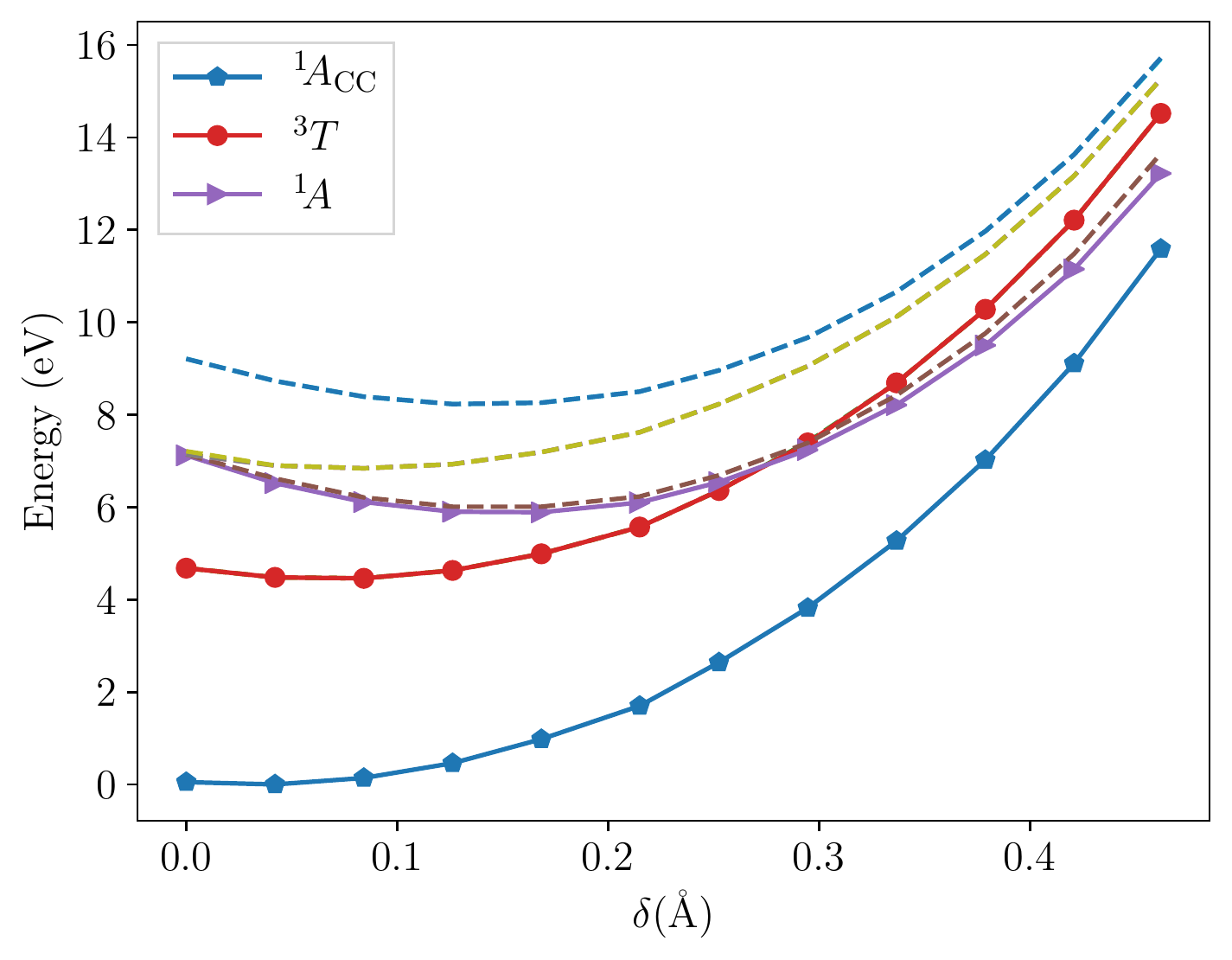}
  \caption{%
    Configuration curve along the phonon $A_{1g}$ mode
    for the excited states of the $F$-center in MgO
    (as shown in Fig~\ref{fig:f_in_mgo}).
    The $^1A_\mathrm{CC}$ curve represents the singlet \gls{uccsd} ground state
    and the upper curves depict the \gls{eeeomccsd} excited states.
    Dashed lines represent \gls{eeeomccsd} states that do not play a role for
    our discussion, but are included for completeness.
    The energies presented are energy differences between the excited
    state energies and the \gls{uccsd} energy.
    The calculation was done for a 4$\times$4$\times$4
    cell containing 127 atoms, 4 active electrons and 64 virtual orbitals.
  }
  \label{fig:configuration-coordinate-mgo}
\end{figure}

Our analysis of the emission process is based on
a Franck-Condon%
~\cite{ElementaryProcFranck1926, ATheoryOfIntCondon1926}
description of the defect.
This is a common approach to treat emission processes
in solids and molecules%
~\cite{ElectronicStruwilson1977%
      , Excited.states.Yuchen2010%
      , doi:10.1063/1.4948245%
      }.
Figure~\ref{fig:configuration-coordinate-mgo}
shows the configuration coordinate diagram along an approximate $A_{1g}$
vibrational mode for the most important \gls{eeeomccsd} excited states and the
\gls{uccsd} ground state singlet $^1A_\textrm{CC}$.
We approximate the atomic displacement along the $A_{1g}$ mode by increasing
the outward displacement of the alkaline earth atoms as depicted in
Fig.~\ref{fig:f_in_mgo}, and keeping all other atomic positions of the
employed 4$\times$4$\times$4 supercell fixed.
The configuration curve has been computed only for MgO but serves as a
qualitatively identical model for CaO and SrO.
Within this picture, the absorption is given by the optically allowed
transition of $^1A_\mathrm{CC} \to {}^1A$ at the ground state
geometry in Fig.~\ref{fig:configuration-coordinate-mgo}.
Taking into account the Franck-Condon approximation, once the
$F$-center is in the excited singlet state, a relaxation of the atoms
along the $^1A_{1g}$ vibrational mode sets off
which could induce a crossing in the configuration curve with the excited
triplet state $^3T$.
Luminescence is then achieved through the transition $^3T \to {}^1A_\mathrm{CC}$.
From the above discussion and the fact that the minimum
of the $^3T$ state is close to the minimum of the ground state,
we conclude that the absorption and emission energies can therefore
be well approximated using the energy differences computed in the
equilibrium structure of the electronic ground state for the $F$-center.

\subsection{MgO}

\begin{table}[t]
  \caption{
    Obtained results from this work
    for the absorption and emission energies of the $F$-centers in MgO, CaO
    and SrO. The \gls{eeeomccsd} results are extrapolated to the complete
    basis set and
    infinite supercell size limit in order to allow for a direct comparison between
    theory and experiment.
    The $GW$ gaps do not correspond to optical excitation energies but are
    included for comparison.
    All energies are in eV units.
  }
  \label{tab:tdl-results}
  \begin{tabular}{llll}
  \toprule
  System & Method & Absorption & Emission \\
  \midrule
  MgO & \gls{eeeomccsd}                                       & 5.28   & 3.66  \\
      & Exp.~\cite{ELECTRONIC.STRUWILSON1976}                 & 5.0    & 2.4   \\
      & QMC.~\cite{PointDefectOperteki2013}                   & 5.0(1) & 3.8(1)\\
      & CASPT2~\cite{OnTheAccuratesousa2001}                  & 5.44   & 4.09   \\
      & $G_0W_0$@LDA0-BSE.~\cite{First.PrincipleRinke.2012}   & 4.95   & 3.4   \\
      & $G_0W_0$@LDA0~\cite{First.PrincipleRinke.2012}        & 5.4    &       \\
      & $G_0W_0$@PBE~\cite{PredictionOfOtosoni2012}           & 4.48   &       \\
      & $GW_0$@PBE~\cite{PredictionOfOtosoni2012}             & 4.71   &       \\
      & $GW$@PBE~\cite{PredictionOfOtosoni2012}               & 5.20   &       \\
  \midrule
  CaO & \gls{eeeomccsd}                                    & 3.13   & 1.93 \\
      & Exp.~\cite{LuminescenceSpbates1974,%
                   HighTemperaturbates1975}                & 3.02   & 1.93 \\
      & Exp.~\cite{ELECTRONIC.STRUWILSON1976}              & 3.1    & 2.05 -- 2.01\\
      & \gls{tddft}@B3LYP~\cite{OpticalAbsorptcarras2006}  & 3.52   & 2.1  \\
      & $G_0W_0$@PBE~\cite{PredictionOfOtosoni2012}        & 3.20   &      \\
      & $GW_0$@PBE~\cite{PredictionOfOtosoni2012}          & 3.53   &      \\
      & $GW$@PBE~\cite{PredictionOfOtosoni2012}            & 3.87   &      \\

  \midrule
  SrO & \gls{eeeomccsd}                      & 2.34 & 1.2\\
      & Exp.\cite{ELECTRONIC.STRUWILSON1976} & 2.4  &    \\
  \bottomrule
  \end{tabular}
\end{table}

The $F$-center in MgO was first discovered by
Wertz et al.~\cite{ElectronSpinRwertz1957} in its positively
charged variant ($F^+$-center) by electron spin resonance measurements,
showing a strong localization of the electrons in oxygen vacancies.
A host of experimental results followed and with it a better
understanding of the absorption and luminescence
mechanisms%
~\cite{ IrradiationDamclarke1957%
      , AnionVacancyChender1980%
      , Theory.of.electVail.1990%
      }.
Experimental and theoretical studies have shown that
the Mg atoms relax in an outward direction from the
vacancy~\cite{ElectronNucleaunruh1967, LatticeDistorthallib1973}.
By using a semi-empirical model, Kemp and Neeley~\cite{WaveFunctionsKemp1963}
predicted an optical absorption energy
of 4.73~eV in good agreement with experimental findings
of 4.95~eV~\cite{IrradiationDamclarke1957, TheTemperaturehender1968}.
The luminescence band of the $F^+$ center was measured at around
3.15~eV~\cite{LuminescenceOfchen1969} while for the $F^0$ center
a luminescence of 2.4~eV was predicted from temperature dependent
measurements of the absorption spectrum in conjunction with a simplified
Huang-Rhys model approach~\cite{TheTemperaturehender1968}.

Using \gls{eeeomccsd} in combination with the outlined extrapolation techniques
yields an absorption and emission energy of 5.2~eV and 3.66~eV, respectively.
Previous many-body ab initio calculations
using $GW$-\gls{bse}~\cite{First.PrincipleRinke.2012},
quantum Monte Carlo~\cite{PointDefectOperteki2013} methods and
CASPT2~\cite{OnTheAccuratesousa2001} agree
with our results for both absorption and emission to within about 0.4~eV as
summarized in Table~\ref{tab:tdl-results}. The calculated absorption energies
are in good agreement with experimental measurements of 5.0~eV.
We note, however, that the $GW$ results (excluding the exciton
binding energy) obtained for different levels of self-consistency and DFT
references exhibit a significant variance ranging from 4.48~eV to 5.4~eV.
Consequently, $GW$-\gls{bse} absorption energies are
strongly dependent on the DFT reference.
Furthermore, we stress that a direct comparison of the computed emission energies
between the quantum chemical approaches (\gls{eeeomccsd} and CASPT2)
and QMC or $GW$ is complicated by the fact that the latter approaches do not
consider the emission process of the de-excitation from the excited triplet states.
Instead, the emission energies computed using QMC and $GW$-\gls{bse} correspond to the
decay from the excited singlet state in its relaxed geometry along the $A_{1g}$ mode.
Nonetheless, from the results shown in
Fig.~\ref{fig:configuration-coordinate-mgo}, we conclude that these different emission
energies are expected to agree to within the errors
made by other approximations.

The measured experimental emission at
2.4~eV~\cite{Photoluminescenedel1979} and its interpretation
is the topic of an ongoing debate.
Initially, this peak has been attributed to the $F$-center and
common interpretations have ranged from a singlet-singlet transition to a
$^3T_{1u} \to {}^1A_{1g}$ transition~\cite{ElectronicStruwood1975,ELECTRONIC.STRUWILSON1976}.
However, it was first suggested by Edel et
al.~\cite{Photoluminescenedel1979,ColouirCentresedel,Photoluminescenedel1982}
that this band results from a recombination process similar to
recombination processes in semiconductors.  Edel and coworkers argue
that the three-electron vacancy $F^{-}$ recombines with the
$F^{+}$-center.
Rinke et al.~\cite{First.PrincipleRinke.2012} have suggested that
the 2.4~eV emission is produced when electrons in
the defect orbitals recombine with the valence holes that can be
produced by intense UV light irradiation. 
The creation of these holes is possibly also
related to the concentration of H$^-$ impurities that are commonly
present in MgO samples, especially when these have been
thermochemically reduced%
~\cite{ DiffusionOfDegonzal1982%
      , ChargeAndMasschen1983%
      , LuminescenceInjeffri1982%
      , LuminescenceOfrosenb1989%
      , LuminescenceFrsummer1983%
      }.
The presence of H$^-$ impurities in MgO could account for the
long-lived luminescence through a hopping mechanism of the electrons
from H$^-$ to H$^-$ impurities until they encounter an $F$-center.
However, it is not immediately clear from the ab initio calculations
thus far if these states are orbital and spin triplets or otherwise as
has been proposed in experimental evidence and symmetry arguments%
~\cite{ELECTRONIC.STRUWILSON1976}.  It has been noticed, however, that
the strength of the 2.4~eV band is temperature dependent as well as
$F$-center and H$^-$ concentration
dependent~\cite{LuminescenceFrsummer1983}.
Typically, neutron irradiation produces mainly $F^+$-centers
while electron irradiation or additive colorization induces mainly
$F$-centers~\cite{Photoluminescenedel1979}.
Rinke et al. argue that given the fact that the position of the
absorption band for the $F$ and $F^+$ centers are almost identical,
it is to be expected that this is also the case for the emission.
Even though similar luminescence peaks for these
centers have been predicted in Ref.~\cite{First.PrincipleRinke.2012}, no
substructure in the emission band can be observed experimentally (unlike in the
absorption band).
Here, we propose a different interpretation of this observation.
We suggest that the $F$-center does not in fact luminesce.
Indeed, modern theoretical computations seem to agree on the fact that
the 2.4~eV band does not belong to the $F$-center luminescence
process. We stress that all theoretical results for the emission energy
summarized in Table~\ref{tab:tdl-results} range from 3.4~eV to 4.09~eV.
Moreover, there is a strong photoconversion from $F$
into $F^+$-centers~\cite{FFcenterConvkapper1972},
suggesting that before the $F$-center has a chance to luminesce,
a conversion into $F^+$ happens followed by an absorption of
the $F^+$-center since the absorption band for it
is similar to the $F$-band.
Our calculations show that the excitation energy for the singlet
state in the $F$-center of MgO converges very slowly with respect to the system
size, indicating that the optically excited state is significantly more
delocalized than the ground state.  This could make a photoconversion into $F^+$
significantly more likely and therefore corroborates our interpretation.

\subsection{CaO and SrO} 

Historically, one of the best studied $F$-centers
in the alkaline earth oxides is the one in CaO%
~\cite{ElectronicStruwood1975}.
The identification of the $F$-center's charged state
is made easier by the fact that, unlike for MgO, the absorption band
is different for the $F$ and $F^+$ centers.
Furthermore, we note that the lattice constant of CaO is significantly
larger than for MgO, which leads to a reduced confinement of the trapped
charges and shifts the absorption band to lower energies.
Early theoretical and experimental investigations
have interpreted the 2.0~eV emission band to be a transition
from a spin and orbital triplet $^3T_{1u}$ into
the ground state singlet $^1A_{1g}$%
~\cite{CouplageJahnTedel1974, LuminescenceSpbates1974, ElectronicStruwood1975}.
However, a $^1T_{1u} \to {}^1A_{1g}$ transition is also possible
at a slightly higher energy.
In general, the CaO luminescence mechanism has been found to be a combination
of a singlet-singlet and a triplet-singlet transition
which are activated at different
temperatures~\cite{ LuminescenceSpbates1974%
                  , HighTemperaturbates1975%
                  }.
Since the excited triplet state lies slightly below in energy
from the excited singlet state, there is a population conversion
at temperatures of around 600~K. Namely, at low temperatures up to
300~K one measures a transition at around 1.98~eV, whereas
as the temperature increases the excited singlet gets populated
and a much more rapid luminescence gets gradually triggered
at around 2~eV%
~\cite{ LuminescenceSpbates1974%
      , HighTemperaturbates1975%
      }.

Using \gls{eeeomccsd} in combination with the outlined extrapolation techniques
yields an absorption and emission energy of 3.13~eV and 1.93~eV for the
$F$-center in CaO, respectively.  To the best of our knowledge only one TD-DFT
result can be found in literature for this system, predicting an absorption and
emission energy of 3.52~eV and 2.1~eV, respectively.
Table~\ref{tab:tdl-results} also summarizes two different experimental
estimates, showing that the \gls{eeeomccsd} and  \gls{tddft}@B3LYP calculations
agree with experiment to within 0.1~eV and 0.5~eV, respectively.  We note again
that $GW$ results for the absorption
energy obtained for different levels of self-consistency shows a significant
variance ranging from 3.2~eV to 3.87~eV and can not be compared
directly to experiment due to the neglect of the exciton binding energy.
We note that our quantum chemical results
have been obtained using periodic boundary conditions,
whereas previous calculations have been carried out using a cluster model approach%
~\cite{OnTheAccuratesousa2001, OpticalAbsorptcarras2006}.

Finally, we turn to the discussion of the $F$-center in SrO. This system exhibits
an even larger lattice constant
and the absorption and emission energies are shifted to even lower energies
compared to MgO and CaO.
However, the $F$-center in SrO is qualitatively
very similar to the CaO case, and the agreement of \gls{eeeomccsd}
in both cases with experimental values is excellent.
To the best of our knowledge, there exist only experimental estimates of the
absorption energy with about 2.4~eV,
whereas no measurements for the emission band are known to the
authors.
We report the results for the
singlet-triplet absorption $^1A_\mathrm{CC}\to {}^3T$
and triplet-singlet emission $^3T \to  {}^1A$
in the infinite supercell size limit in Table~\ref{tab:tdl-results}. We hope that this
prediction will be verified experimentally in the future.

\section{Conclusions}

In this work we have presented a novel implementation of
the \gls{uccsd} and \gls{eeeomccsd} methods
for periodic systems using a plane wave basis set and applied them to the
$F$-center
in the alkaline earth oxides MgO, CaO, and SrO.
The implementation was tested on molecular systems, and we
have verified it by comparing against well established quantum chemistry
codes for a number of molecular and atomic systems.
Convergence of calculated excitation energies with respect to
the basis set and size of the simulation cell is crucial for reliable
predictions in periodic systems.
We have presented a framework to obtain basis-set and finite-size corrected
excitation energies by freezing the number of occupied
orbitals in a controlled fashion and extrapolating to the complete basis set and
infinite system size limit.

We have calculated \gls{eeeomccsd} absorption and emission energies of the
$F$-center in
MgO, CaO, and SrO, accounting for finite basis set and system size errors using
extrapolation techniques.
The obtained results are in good agreement with previous
calculations (where available)%
~\cite{ First.PrincipleRinke.2012%
      , PointDefectOperteki2013%
      , OpticalAbsorptcarras2006%
      , PredictionOfOtosoni2012%
      , OnTheAccuratesousa2001%
      }
and with experimental data~\cite{ELECTRONIC.STRUWILSON1976}.
In addition, a prediction for the emission band of the $F$-center in SrO
has been made.
Furthermore, we provide additional evidence for the
assignment of the 2.4~eV band in MgO crystals to recombination processes,
and we propose a new interpretation of previous results by
suggesting that the $F$-center in MgO does not luminesce.
However, further work is needed to clarify the
nature of these transitions. 

The achieved level of accuracy for the calculated \gls{eeeomccsd} absorption
and emission energies shows that this method
has the potential to significantly expand the scope of currently available
ab initio techniques for the study of defects.
However, further improvements for the corrections to the finite basis and system size
errors are urgently needed to allow for a more extensive and detailed study of
defects in solids on the level of \gls{eeeomccsd} theory.

\section{Acknowledgements}
The authors thankfully acknowledge support and funding from the European
Research Council (ERC) under the European Unions
Horizon 2020 research and innovation program (Grant Agreement No 715594).
The computational results presented have been achieved in part using
the Vienna Scientific Cluster (VSC).

\bibliography{main.bib}

\end{document}